\documentclass{article}

\usepackage{arxiv}

\usepackage[utf8]{inputenc} 
\usepackage[T1]{fontenc}    
\usepackage{hyperref}       
\usepackage{url}            
\usepackage{booktabs}       
\usepackage{amsfonts}       
\usepackage{nicefrac}       
\usepackage{microtype}      
\usepackage{lipsum}		
\usepackage{graphicx}
\usepackage{natbib}
\usepackage{doi}

\title{Understanding the Kerala Floods of 2018:\\Role of Mixed Rossby-Gravity Waves}

\date{} 					

\author{ \href{https://orcid.org/0000-0003-3200-8624}{\includegraphics[scale=0.06]{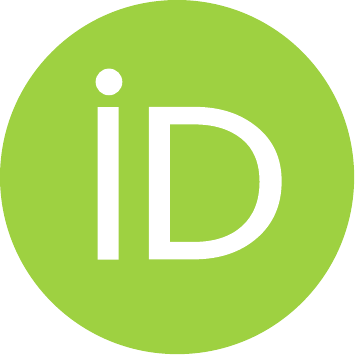}\hspace{1mm}Kiran S. R.}\thanks{Corresponding author \emph{(kiransreekumarr@gmail.com, ce\_kiran@cpt.ac.in)}} \\
	Lecturer in Civil Engineering, Central Polytechnic College\\
	Department of Technical Education\\
	Thiruvananthapuram - 695013\\
	Kerala, India\\
	\vspace{0.25cm}\\
	\emph{\small Submitted on $17^{th}$ March 2021}\\
}

\renewcommand{\headeright}{Kiran S. R.}
\renewcommand{\undertitle}{}
\renewcommand{\shorttitle}{Understanding the Kerala Floods of 2018: Role of Mixed Rossby-Gravity Waves}

\hypersetup{
pdftitle={Understanding the Kerala Floods of 2018: Role of Mixed Rossby-Gravity Waves},
pdfsubject={q-bio.NC, q-bio.QM},
pdfauthor={Kiran S. R.},
pdfkeywords={Kerala, Floods, Monsoon, Intertropical Convergence Zone, Mixed Rossby-Gravity Waves},
}

\begin{document}
\maketitle

\begin{abstract}
	Kerala, the south-west coastal state of India, was ravaged by a series of floods during the South-West Monsoon of 2018. The season was marked by severely anomalous rainfall trends, with upto 150 mm of departures from the mean daily precipitation in the northern districts of the State. Although, there were many studies about the hydrogeological factors which aggravated the floods in Kerala, no attempt was made to delve into the physics which actually resulted in anomalous precipitation during the year. This study intends to document the dynamical phenomenon which caused the Kerala Floods of 2018. The westward propagating convectively-coupled Mixed Rossby-Gravity (MRG) waves were excited by the synoptic disturbances of the tropical Pacific at the pressure level of 700 hPa, during the Indian Monsoon of 2018. They travelled across the Indian Ocean in two significant modes -- a predominant slow moving wave of 20-40 days period (as Madden-Julian Oscillation) and a secondary faster wave of 5-8 days period. They are characterised by vertical phase propagation to the upper troposphere, a precursor to deep convection and intense precipitation. Further, the propagation of these waves through a medium enhances its relative vorticity and the gyres or circulations thus formed are symmetrical about the equator. Consequently, the meanders in the wind field and widening of the Intertropical Convergence Zone were observed. The MRG waves, especially the slow mode induced divergence in the wind field, which fueled convection in tropics and brought very heavy rainfall to the State of Kerala in 2018.
\end{abstract}


\section{Introduction}
Climate of any place is the product of inextricable intertwining of the atmosphere, ocean, land, ice and biology. Climate change has become a buzzword today and is marked by frequent occurrences of extreme weather events -- severe floods and droughts \citep{anke2008,magnus2019}. It is observed that the frequency of occurrences of short-duration extreme precipitation events have increased in different parts of the world, including India.

Kerala, the State on south-west coast of India, experienced torrential rainfall and devastating floods in the year 2018 during Boreal Summer. The South-West Monsoon season observed an exceptionally high precipitation in the State with over 15\%, 18\% and 164\% above normal rainfall during the months of June, July and August respectively \citep{imd2018}. Kerala received a total of 821 mm of precipitation only in the month of August, which is the highest recorded rainfall for the month after 1931. A series of unprecedented deluge events ensued the heavy downpour and 13 out of the 14 districts in the State witnessed floods of disastrous proportions. Further, the State witnessed recurrent flood events, during the same season, in the subsequent years 2019 and 2020.

\citet{vimalmishra2019} observed that there is no increase in mean and extreme precipitation in Kerala over the last six decades and the extreme rainfall event during August 2018 is driven by anomalous atmospheric conditions due to climate variability rather than anthropogenic factors. They concluded that the severity of the Kerala flood of 2018 was exacerbated by several other factors such as land use/ land cover change, antecedent hydrologic conditions, reservoir storage and operations, encroachment of flood plains, etc. \citet{preet2020} attempted to perform inundation mapping of Kerala, to identify the flood-affected regions in the State. \citet{sudheer2019} analyzed the role of dams and reservoir operations, on the flood of August 2018. They deduced that the role of water, released from the major reservoirs in the Periyar River Basin, to have resulted in the flood havoc was minimal. \citet{hunt2020}, using Weather Research and Forecasting (WRF) model in tandem with a hydrological model, performed a broad analysis to underscore climate change and concluded that the precipitation may be 18\% heavier for the pre-industrial period (due to recent weakening of monsoon low-pressure systems), and 36\% heavier for a futuristic climate of 2100 (due to moistening of the tropical troposphere).

Previous studies, about the Kerala Floods of 2018, have neither documented the observations nor attempted to analyze deeper into the physical phenomenon responsible for the disaster. Hence, any investigation into the anomalous climate dynamics, which triggered the deadly flood events in the State, is yet to pull off. Further, the recurrence of severe flood events in Kerala in the recent years arose the need to identify this anomaly at the earliest, which became the prime motivation for this study.

\section{Data \& Methodology}
\label{sec:dataandmethodology}
The daily rainfall data over the Indian Ocean region is obtained from Tropical Rainfall Measuring Mission (TRMM) at a spatial resolution of $0.25^{o}$ $\times$ $0.25^{o}$, for the period from 2011 to 2018. It gives the daily accumulated precipitation (combined Microwave-IR) estimate with gauge calibration performed over land in millimetres. The globally gridded data of zonal and meridional winds at a spatial resolution of $2.5^{o}$ $\times$ $2.5^{o}$ from 2011 to 2018 is obtained from NCEP/NCAR Reanalysis 1, which uses data assimilation to incorporate observations and numerical weather prediction (NWP) model output, for accurate representation of the state of Earth's atmosphere. This gives horizontal wind (zonal and meridional) data (in metres per second) for different vertical pressure levels such as 1000, 925, 850, 700, 600, 500, 400, 300, 250 and 200 in hectopascals. In order to analyse the probable effects of convection, omega or temporal derivative of pressure (in pascals per second), which is an estimate of vertical velocity for an atmosphere in pressure coordinate system, is also obtained for different vertical levels, as mentioned above, from NCEP/NCAR Reanalysis.

In this study, the daily departures from the mean of atmospheric variables during 2018 are determined to quantify the gravity of anomalous precipitation in Kerala. This is estimated by deducting the average of daily data of the atmospheric variables from 2011 to 2017 (7year daily mean) from the daily data for the year 2018. In order to compute the width of Intertropical Convergence Zone (ITCZ), a threshold rainfall of 50mm/day is chosen and the latitudinal extend over which the spatial continuous rainfall outstrip the threshold along the desired meridian is deemed to be the apparent width of ITCZ. Computation of the relative vorticity involves determination of spatial derivatives of winds, which is performed by means of simple finite-backward differencing method, as executed by \citet{kiran2017} to determine the curl of wind stress in Andaman Sea. Hovmöller diagrams, which depict the temporal variation of atmospheric parameters in the zonal or meridional direction, are used to identify and explain atmospheric waves and its characteristics. Fast Fourier Transform (FFT) is performed on time series of atmospheric variables or its anomalies for the ease of representation in frequency domain. The dominant harmonics, thus identified, are filtered out using Band-Pass Filters (BPF) for further analysis. 

\section{Results \& Discussions}

\subsection{Rainfall over Kerala in 2018}
\begin{figure}
	\centering
	\includegraphics[width=0.9\linewidth]{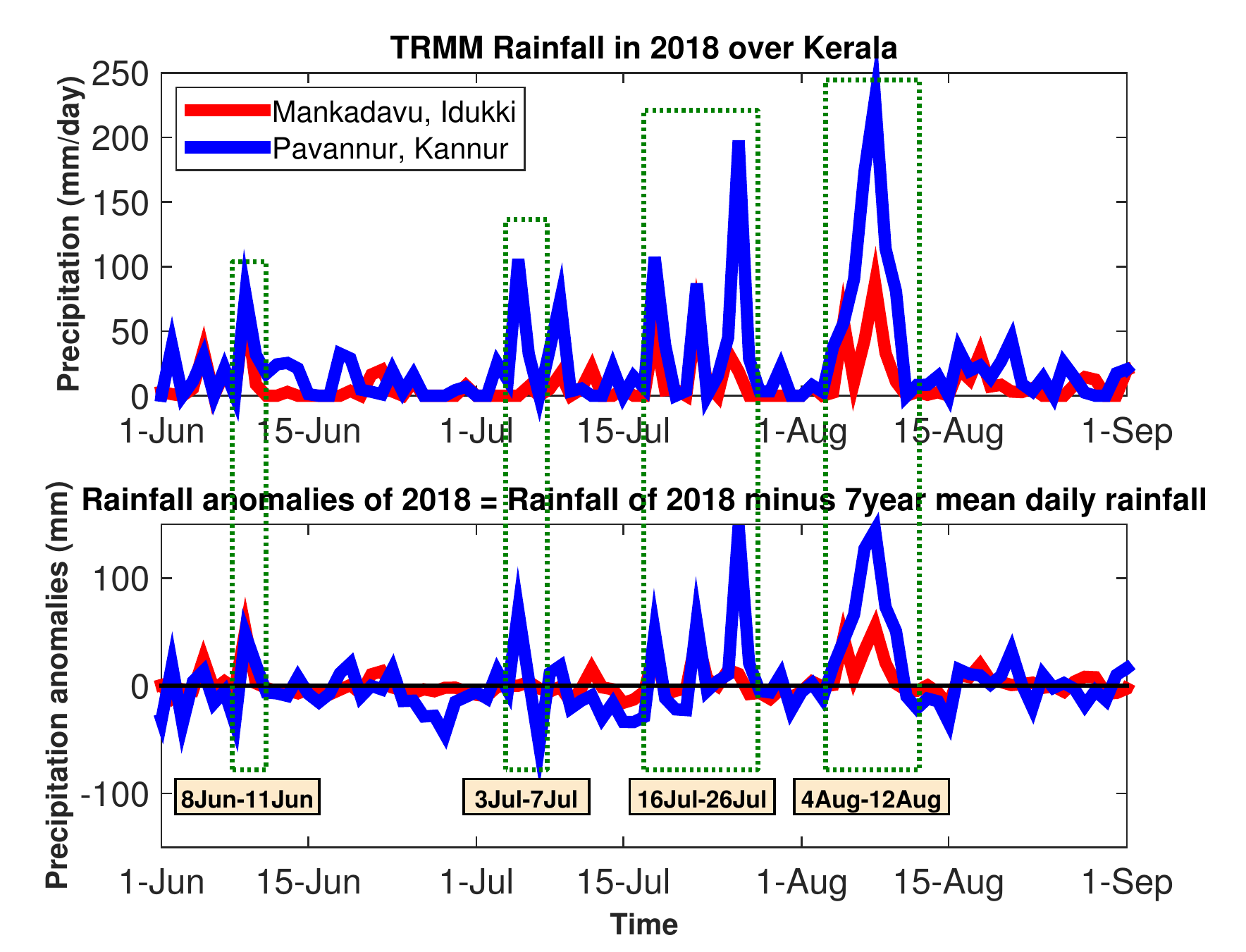}
	\caption{(top) TRMM daily accumulated precipitation (combined Microwave-IR) estimate with gauge calibration performed over land (in millimetres) at Mankadavu (Idukki District) and Pavannur (Kannur district) of Kerala in 2018. (bottom) Daily rainfall anomalies (in millimetres) of 2018, measured with respect to 7 year mean at Mankadavu and Pavannur.}
	\label{fig:fig1}
\end{figure}

The Kerala Flood of 2018 was the result of intense precipitation during the South-West Monsoon over the Indian subcontinent during the months of June to August. The daily rainfall during the period at Mankadavu (Idukki, Central District of Kerala) and Pavannur (Kannur, Northern District of Kerala), as obtained from TRMM dataset, is shown in Figure \ref{fig:fig1} (top). Pavannur leads in terms of rainfall intensity compared to Mankadavu, but both the locations captured the major trends in precipitation. It is observed that daily rainfall in excess of 50mm, occurred at both the locations during the periods $8^{th}$ - $11^{th}$ June, $3^{rd}$ - $7^{th}$ July, $16^{th}$ - $26^{th}$ July and $4^{th}$ - $12^{th}$ August, although Mankadavu missed the peak rainfall during the first week of July. The above time durations are considered as occurrences of extreme weather events and are designated as T1, T2, T3 and T4 respectively (refer Table ~\ref{tab:table1} to compare the rainfall trends at both the stations). A very heavy rainfall of 200 mm/day and above occurred by the end of July and beginning of August, which aggravated the flood events in the State. To quantify the rainfall anomalies, the precipitation data from TRMM for 7 years (2011 to 2017) is daily-averaged and subtracted from the rainfall of 2018 (Fig. \ref{fig:fig1} (bottom) \& Table ~\ref{tab:table1}). The year 2018 saw an extremely anomalous monsoon rainfall during T1, T2, T3 and T4, with excess daily precipitation as high as 150 mm, and hence, this phenomenon requires to be investigated.

\begin{figure}
	\centering
	\hspace*{-2cm}
	\includegraphics[width=1.22\linewidth]{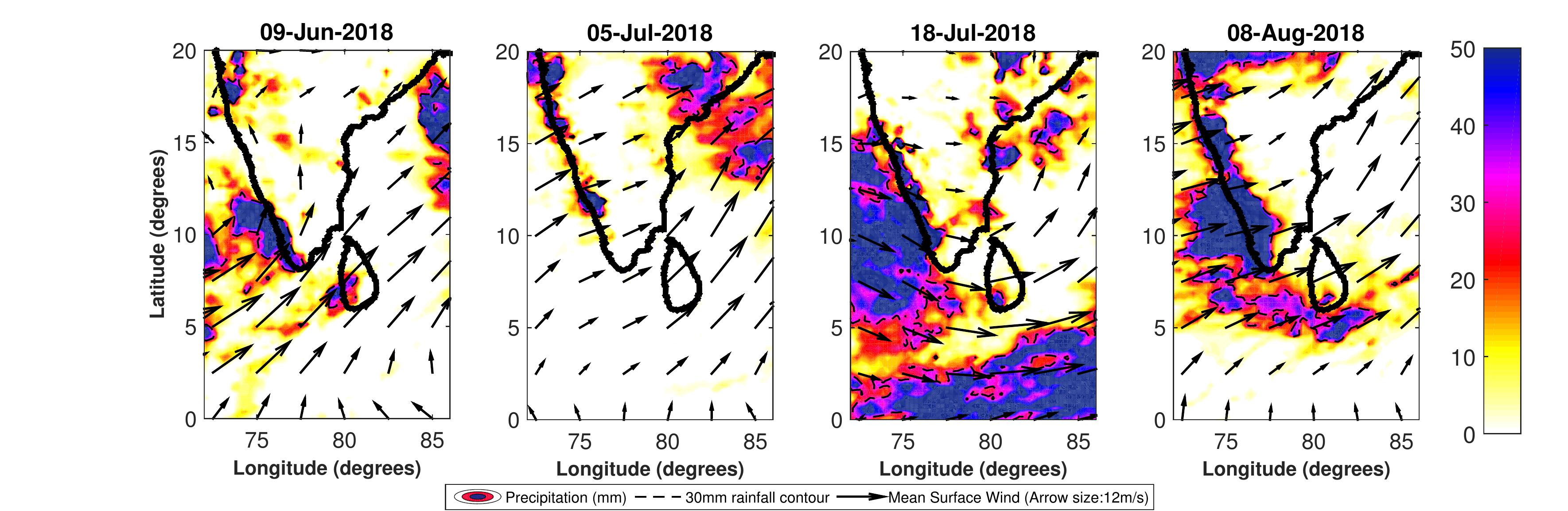}
	\caption{Spatial distribution of TRMM daily accumulated precipitation (in millimetres) overlaid with NCEP/NCAR Reanalysis surface winds (at 1000 hPa) on the days of peak rainfall during T1, T2, T3 and T4.}
	\label{fig:fig2}
\end{figure}

Figure \ref{fig:fig2} shows the spatial distribution of TRMM precipitation on the days of peak rainfall each during T1, T2, T3 and T4 respectively, overlaid with NCEP/NCAR Reanalysis surface wind vectors. On all these days, the entire extend of the state of Kerala received abnormally high precipitation (exceeding 50 mm per day) compared to its neighborhood. It may be observed that the rainfall intensified on the windward side of Western Ghats, with the wind vectors oriented mostly normal to the chain of mountains. This underscores the role of orography in galvanizing Monsoon precipitation. On 5th July 2018, the heavy rainfall is spread only across northern part of the State, which explains why the anomaly during T2 was negligible at Mankadavu. Further, one observes very severe horizontal wind shear and the associated tendency for cyclonic circulation over the region on the days of peak rainfall, especially during T3 and T4. Here, the positive relative vorticity sets high prospects for good rainfall.

\subsection{Analysing the behaviour of ITCZ}
Intertropical Convergence Zone (ITCZ) or the Equatorial trough is an atmospheric system responsible for the large-scale rainfall over tropical oceans. It functions as the ascending limb of the meridional circulation (known as Hadley cell) at lower latitudes of the atmosphere. Indian monsoon is the manifestation of seasonal migration of ITCZ. Further, it is associated with a series of northward propagation of the cloud bands from the equatorial Indian Ocean onto the mainland \citep{riehl1954,charney1969,gadgil2018}.

\begin{figure}
	\centering
	\includegraphics[width=0.9\linewidth]{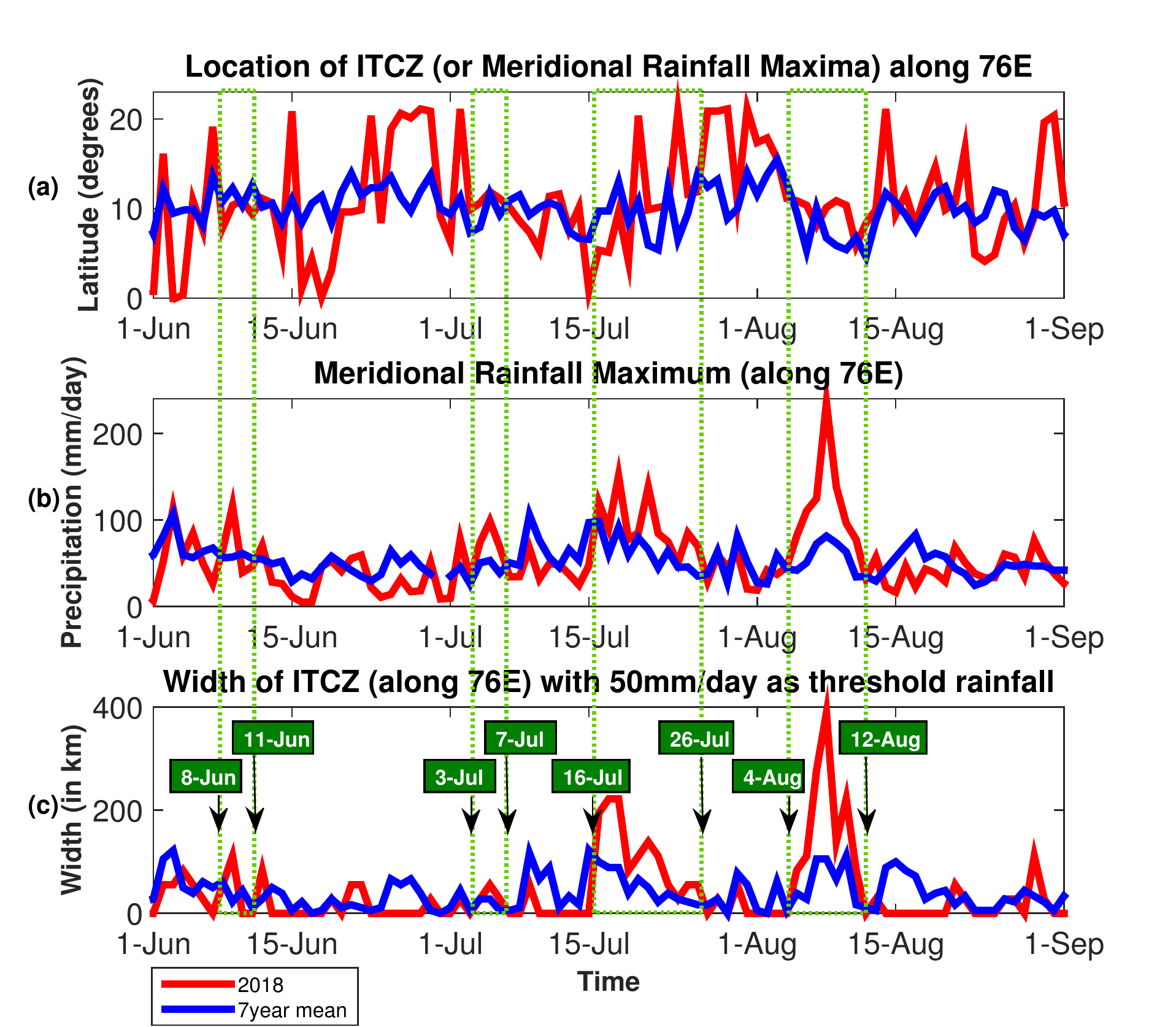}
	\caption{(a) The daily variation of latitudinal position of Intertropical Convergence Zone (ITCZ) along $76^{o}$E. (b) The meridional daily rainfall maximum inside ITCZ in millimeters along $76^{o}$E. (c) The daily variation in the width of ITCZ in kilometres. Red curve indicates the values corresponding to the year 2018 and blue curve indicates the values averaged over 7 years.}
	\label{fig:fig3}
\end{figure}

The location of ITCZ for the given year is chosen as the latitude corresponding to the meridional maximum daily precipitation \citep{byrne2018}. The daily variation of the position of ITCZ along $76^{o}$E, the meridian passing through the State of Kerala, is shown in Figure \ref{fig:fig3} (a), in comparison with the 7 year (2011-2017) daily mean location of the Equatorial trough. During T1, T2, T3 and T4, the ITCZ is positioned over Kerala, although it considerably varies in location and even fell out-of-phase with the mean during T3 and T4. Figure \ref{fig:fig3} (b) is the evolution of meridional rainfall maxima, i.e., the peak rainfall within ITCZ in 2018. One finds the ITCZ rainfall itself is the greatest during T1, T2, T3 and T4, as against the mean. Further, there was a simultaneous increase in the width of ITCZ during T3 and T4 (Fig. \ref{fig:fig3} (c)), where the trough width is obtained as the latitudinal extend about the precipitation maxima along which the daily rainfall exceeds a threshold of 50mm. Here, the widening of ITCZ may be attributed to the perturbations caused to the south-westerly winds, by any local or remote phenomenon \citep{byrne2019}. Hence, ITCZ was approximately positioned over Kerala during the occurrences of extreme rainfall, which widened and delivered heavy precipitation over the State.

\subsection{Understanding the waves and its dynamics}
\begin{figure}
	\centering
	\includegraphics[width=1.05\linewidth]{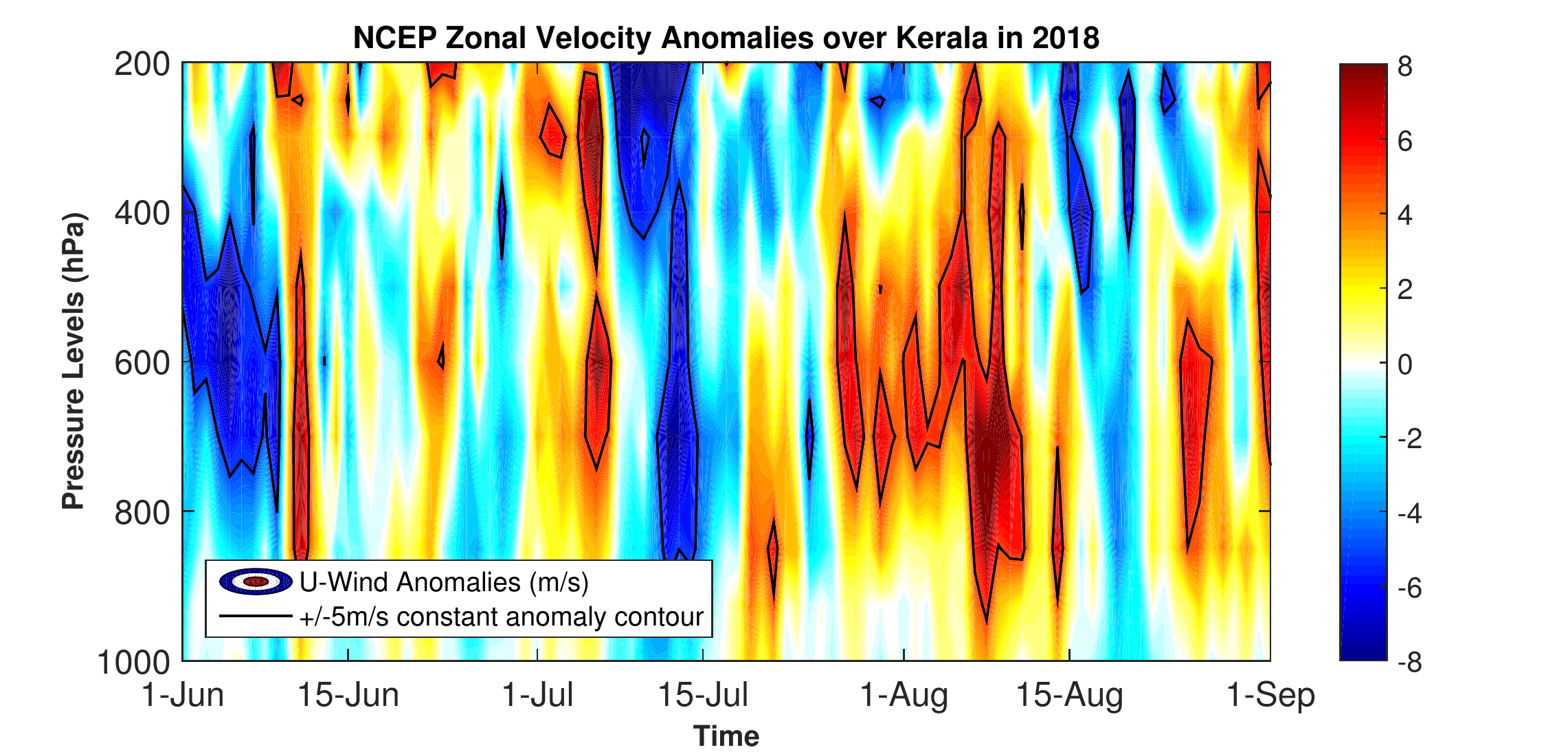}
	\caption{Variation of NCEP/NCAR Reanalysis Zonal Wind Anomalies (in metres per second) with altitude (as pressure levels in hecto pascals) over Kerala (averaged over $75^{o}$E to $77^{o}$E \& $8^{o}$N to $12^{o}$N).}
	\label{fig:fig4}
\end{figure}

Indian Monsoon is said to be strong if the zonal component of the Somali Jet is greater by more than a standard deviation of the long-term mean. The zonal component of the wind is generally stronger during the season, analysing which shall render a better insight into the associated intraseasonal variability \citep{gadgil2018,raghu2007}. NCEP/NCAR zonal wind anomalies over Kerala (averaged over $75^{o}$E to $77^{o}$E \& $8^{o}$N to $12^{o}$N) in 2018, is shown in Figure \ref{fig:fig4}. In comparison to the surface, the westerly winds intensify during T1, T2, T3 and T4 at 800 to 600 hPa pressure levels, while the negative anomalies dominate 400 hPa to 700 hPa during the early weeks of June and July. The occurrence of high positive zonal wind anomalies as high as 8 m/s at an altitude of average pressure 700hPa is concomitant with the periods of anomalous precipitation, i.e., during T1, T2, T3 and T4. The perturbations tend to propagate vertically upwards by the end of July from 700 hPa to upper troposphere and then downwards to the surface. This is characteristic of the vertical propagation of waves in atmosphere. Since, the wind anomalies appear to concentrate near the pressure level of 700 hPa, it shall be chosen as a reference for further investigation.

\subsubsection{Source of perturbation}
\begin{figure}
	\centering
	\hspace*{-1cm}
	\includegraphics[width=1.1\linewidth]{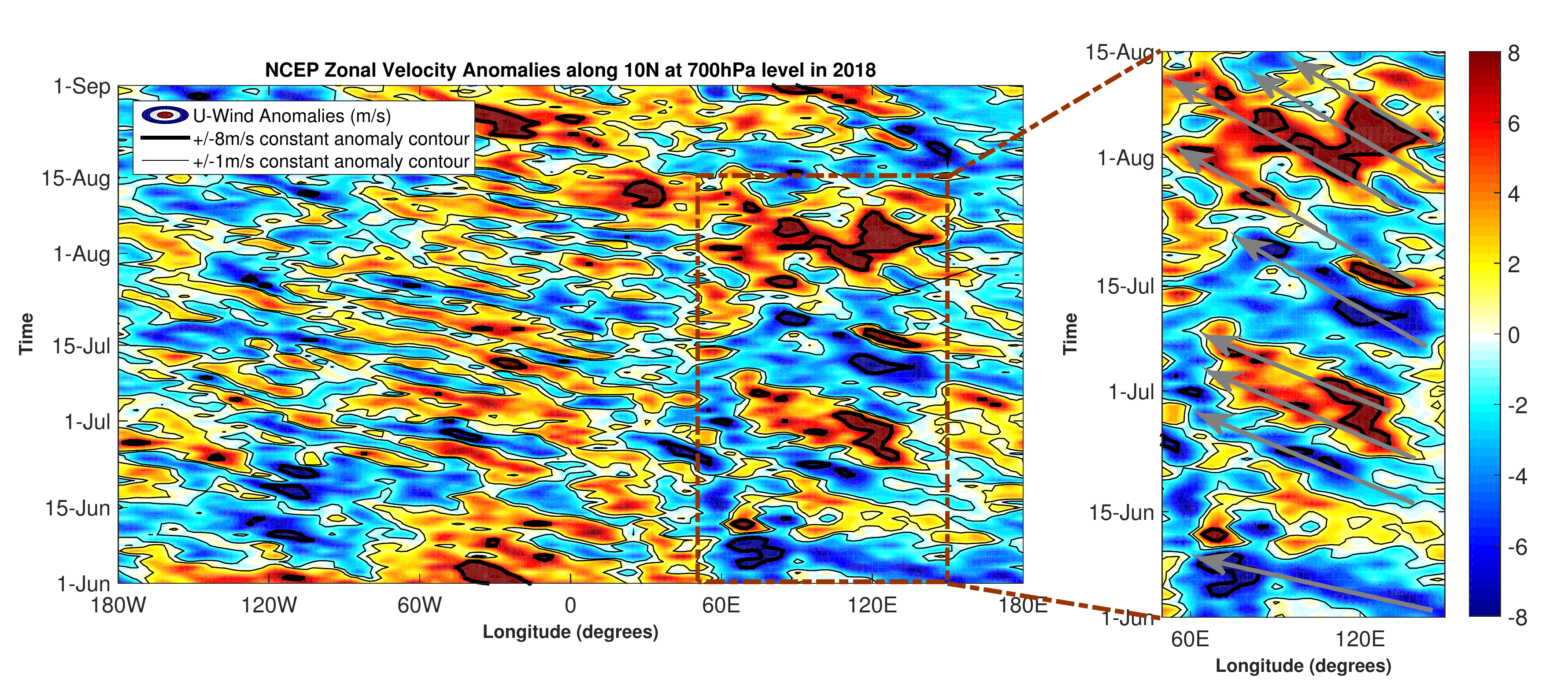}
	\caption{The Hovemöller diagram showing temporal evolution of NCEP/NCAR Reanalysis Zonal Wind Anomalies (in metres per second) at 700hPa pressure level along $10^{o}$N. The segment of the diagram from $50^{o}$E to $150^{o}$E is enlarged for detailed inspection. Grey colored arrows indicate phase propagation of signals in the westward direction.}
	\label{fig:fig5}
\end{figure}

Figure \ref{fig:fig5} gives a neat picture of the evolution of NCEP/NCAR zonal wind anomalies of 2018 at 700 hPa along $10^{o}$N latitude, which passes through Kerala. This Hovemöller diagram exposes westward propagating synoptic disturbances or equatorial waves, which originated from East Atlantic and Pacific Ocean basins, during the months of June to September. They have wavelengths of the range 2500 - 5000 km, and manifest as shallow disturbances (as large wind anomalies at 700-850 hPa levels) embedded in easterly trade winds. These may form as a result of instabilities in the mean flow or due to barotropic instability of the ITCZ \citep{reed1971,ferreira1997,serra2008}, yet the dynamics of initiation of tropical synoptic disturbances hitherto remains a mystery. These westward-propagating signals travelled to eastern end of Equatorial Indian Ocean and perturbed the wind system there. As a result, the zonal wind anomalies over the Indian Ocean basin ($50^{o}$E to $100^{o}$E) intensified.

\subsubsection{Mixed Rossby-Gravity (MRG) Waves}
\label{subsubsec:mrg}
\begin{figure}
	\centering
	\hspace*{-1cm}
	\includegraphics[width=1.1\linewidth]{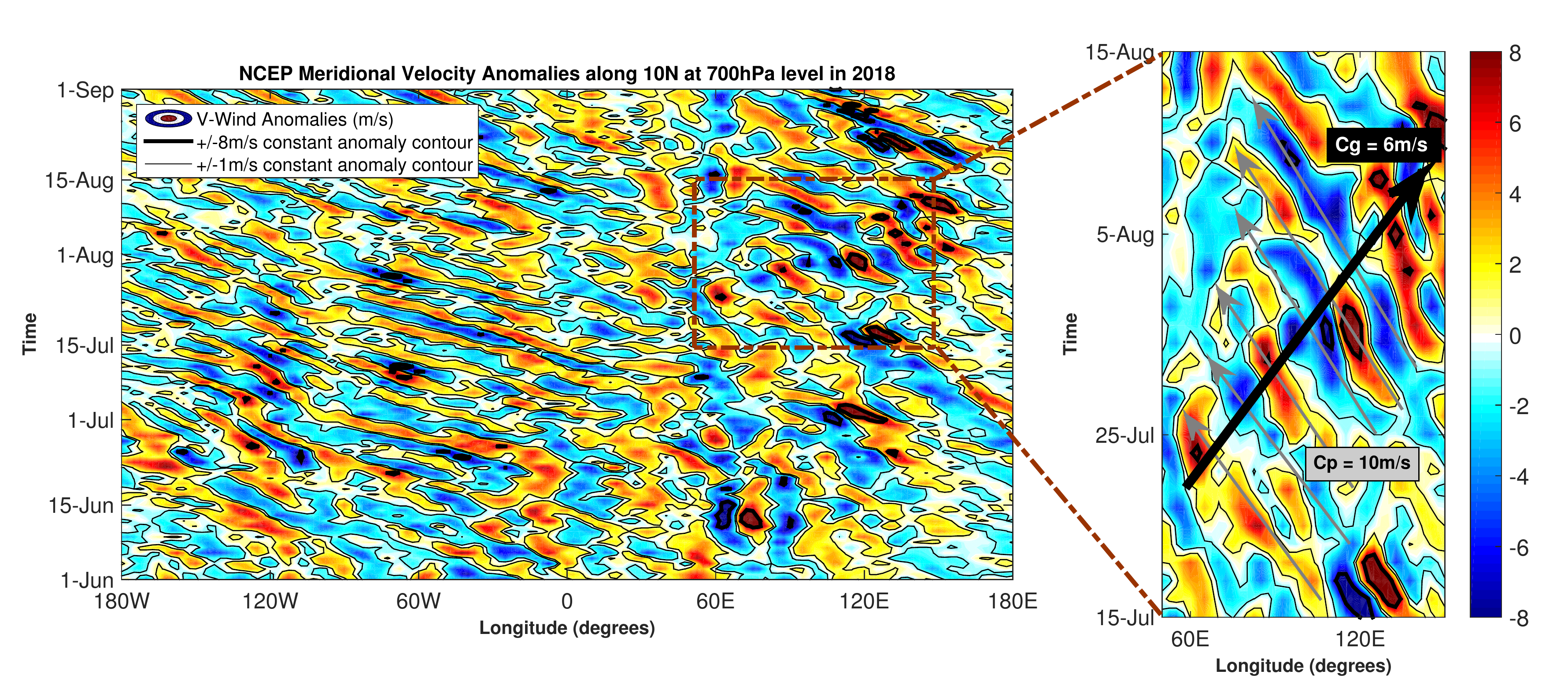}
	\caption{The Hovemöller diagram showing temporal evolution of NCEP/NCAR Reanalysis Meridional Wind Anomalies (in metres per second) at 700hPa pressure level along $10^{o}$N. The segment of the diagram from $50^{o}$E to $150^{o}$E is enlarged for detailed inspection. Grey colored arrows indicate phase propagation of signals in the westward direction and black colored arrow indicates the eastward propagation of the wave group. $C_{p}$ and $C_{g}$ represents the phase and group velocities of waves respectively.}
	\label{fig:fig6}
\end{figure}

The enlarged segment of the Hovemöller diagram (Fig. \ref{fig:fig5}) exposes the intensification of zonal wind anomalies accompanied by westward propagation of signals between 60E and 120E. This reveals synoptic scale waves which originate near $150^{o}$E (western Equatorial Pacific) and propagate westwards till $60^{o}$E (near the coast of Oman), via Bay of Bengal and Arabian Sea. These propagations seem to amplify and coincide exactly with the occurrences of anomalous rainfall in Kerala, especially during T3 and T4. From Figure \ref{fig:fig6}, the NCEP/NCAR meridional wind anomalies over the tropics ($50^{o}$E to $150^{o}$E) support the same events of waves (phase) propagating westward, and the anomalies change sign eastwards (wave group). The phase propagates westward with a speed ($C_{p}$) of 10 m/s, while the wave packets seem to propagate eastward with a group velocity ($C_{g}$) of 6 m/s. The wave propagation characteristics resemble Mixed Rossby-Gravity (MRG) waves, which are westward-propagating equatorially-trapped waves with synoptic time scales. Earlier studies identified lower-tropospheric MRG waves with 3–8 days period in the central Pacific having wavelength of 6000–10000 km, a westward phase speed of 10–20 m/s, and an eastward group speed of nearly 5 m/s \citep{dickinson2002,raupp2005,chen2009,yang2011}. Therefore, MRG waves, triggered by the westward propagating synoptic disturbances of the Pacific, intensified the monsoon rainfall over Kerala as they propagated across the equatorial Indian Ocean region in 2018.

\subsubsection{Characteristics of MRG Waves} 
\begin{figure}
	\centering
	\hspace*{-1cm}
	\includegraphics[width=1.1\linewidth]{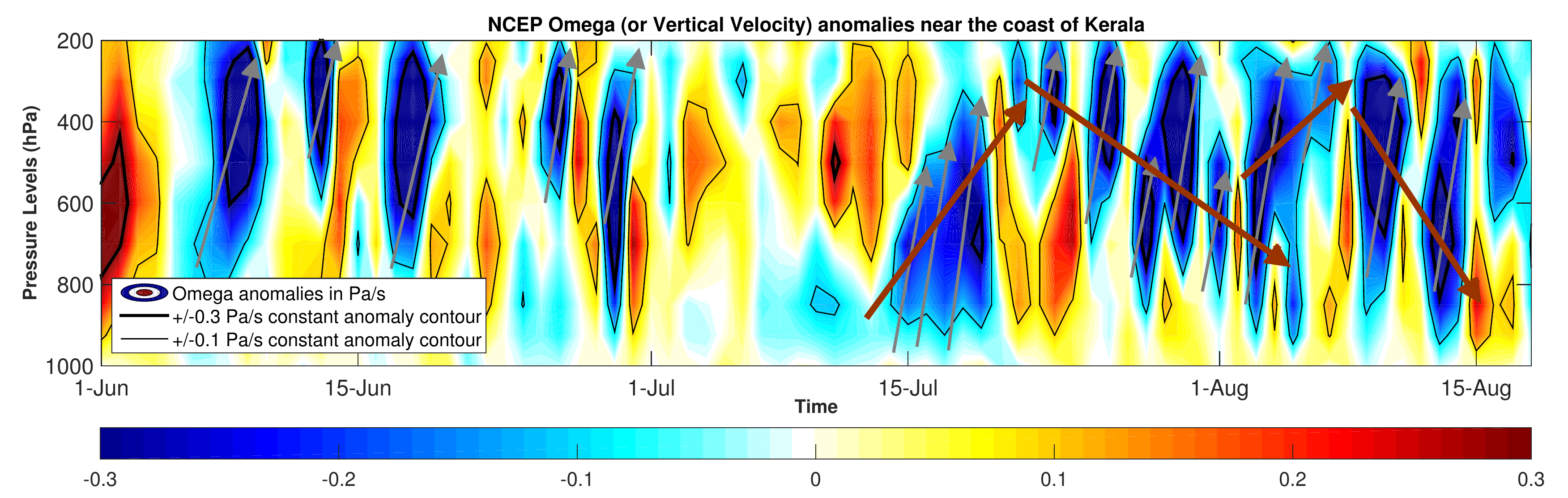}
	\caption{The variation of NCEP/NCAR Reanalysis Omega (indicative of Vertical velocity) Anomalies (in Pa/s) over Kerala (averaged over $75^{o}$E to $77^{o}$E \& $8^{o}$N to $12^{o}$N) in 2018. Negative (positive) values represent rising (sinking) of air parcels. Grey colored arrows indicate upward propagation of the phase of waves and brown colored arrows indicate the propagation of wave group.}
	\label{fig:fig7}
\end{figure}

The NCEP/NCAR omega anomalies of 2018 averaged over $75^{o}$E to $77^{o}$E \& $8^{o}$N to $12^{o}$N (Fig. \ref{fig:fig7}) show clear signs of deep convective activity near the coast of Kerala in the Summer monsoon of 2018, especially during T3 and T4. It elicits an interesting wave character; there occurs upward phase propagation (of negative omega or rising motion of air) from 700 hPa pressure level to upper troposphere which supports deep convection and the consequent intense precipitation is concurrent with the periods of occurrence of anomalous rainfall. Besides, the wave group seem to propagate upward and downward across the height of troposphere, which substantiates the observations discussed on the anomalies of zonal velocity (Fig. \ref{fig:fig4}). The rising motion of the wave group appears to be a precursive event to the anomalous precipitation, while the downward propagation follows it. 

\begin{figure}
	\centering
	\hspace*{-1cm}
	\includegraphics[width=1\linewidth]{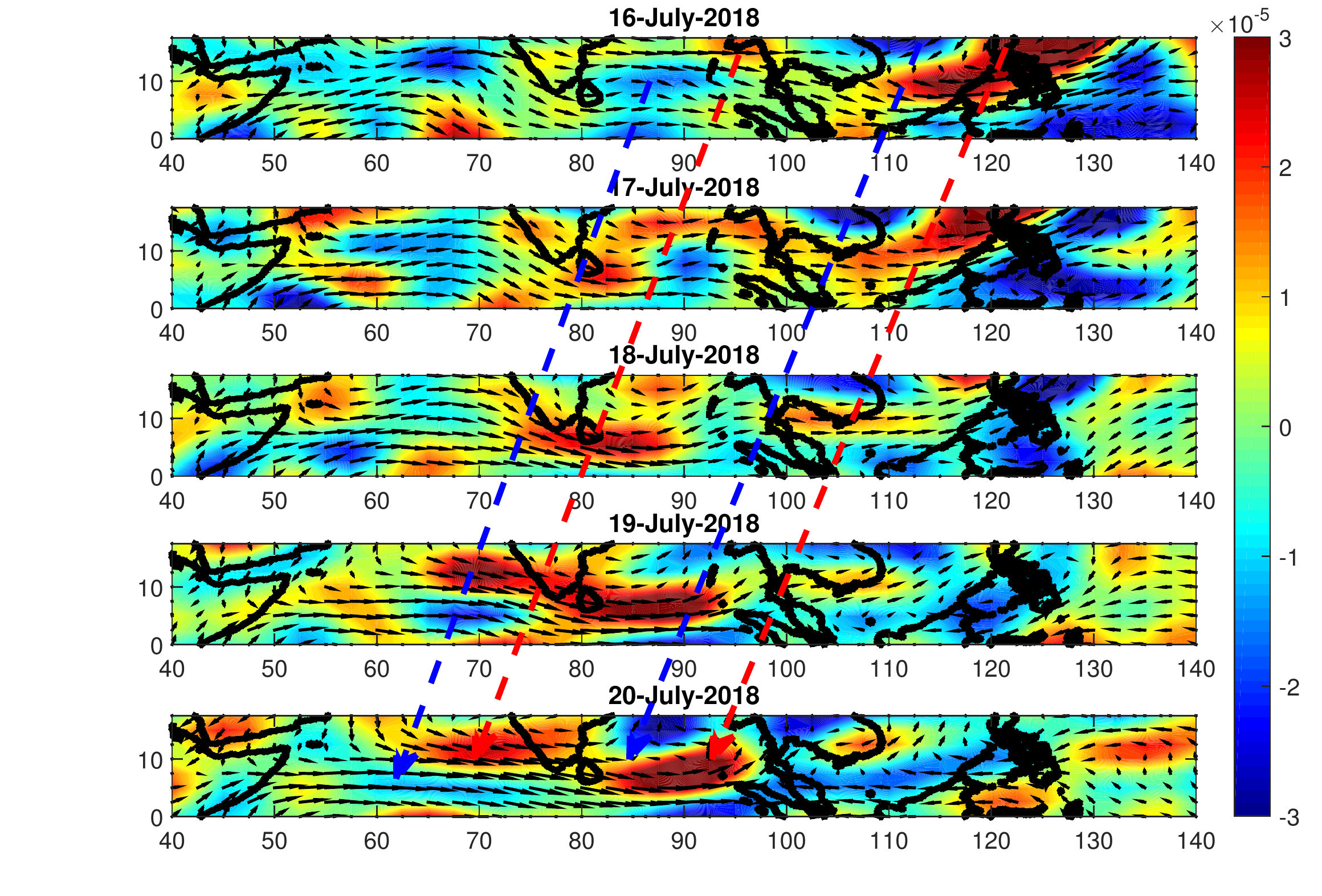}
	\caption{Spatial variation of NCEP/NCAR Reanalysis Relative vorticity Anomalies (in $s^{-1}$) over the Equatorial Indian Ocean at 700 hPa level from $16^{th}$ July to $20^{th}$ July 2018. The dashed red (blue) arrows indicate the westward propagation of cyclonic (anti-cyclonic) vorticity anomalies.}
	\label{fig:fig8}
\end{figure}

It is important to quantify the relative vorticity of the wind field, as waves propagate through the medium. The strong shear in winds during the monsoon of 2018 was already discussed with Figure \ref{fig:fig2}, which shall favour an increase in the flow vorticity. The daily spatial variation of anomalies in the relative vorticity at 700 hPa level overlaid with winds at the same pressure level from $16^{th}$ July to $20^{th}$ July 2018 are shown in Figure \ref{fig:fig8}. The winds near Western Pacific possess very high vorticity of atleast 3 $\times$ $10^{-5}$ $s^{-1}$ on $16^{th}$ July, which tends to shift westward in the subsequent days and intensify over the Bay of Bengal and Arabian Sea on $19^{th}$ July. The wind field appears to possess a meandering or wavy pattern, which may be attributed to the westward propagating MRG waves. 

\subsection{Spectral Analysis}
\begin{figure}
	\centering
	\hspace*{-1cm}
	\includegraphics[width=0.95\linewidth]{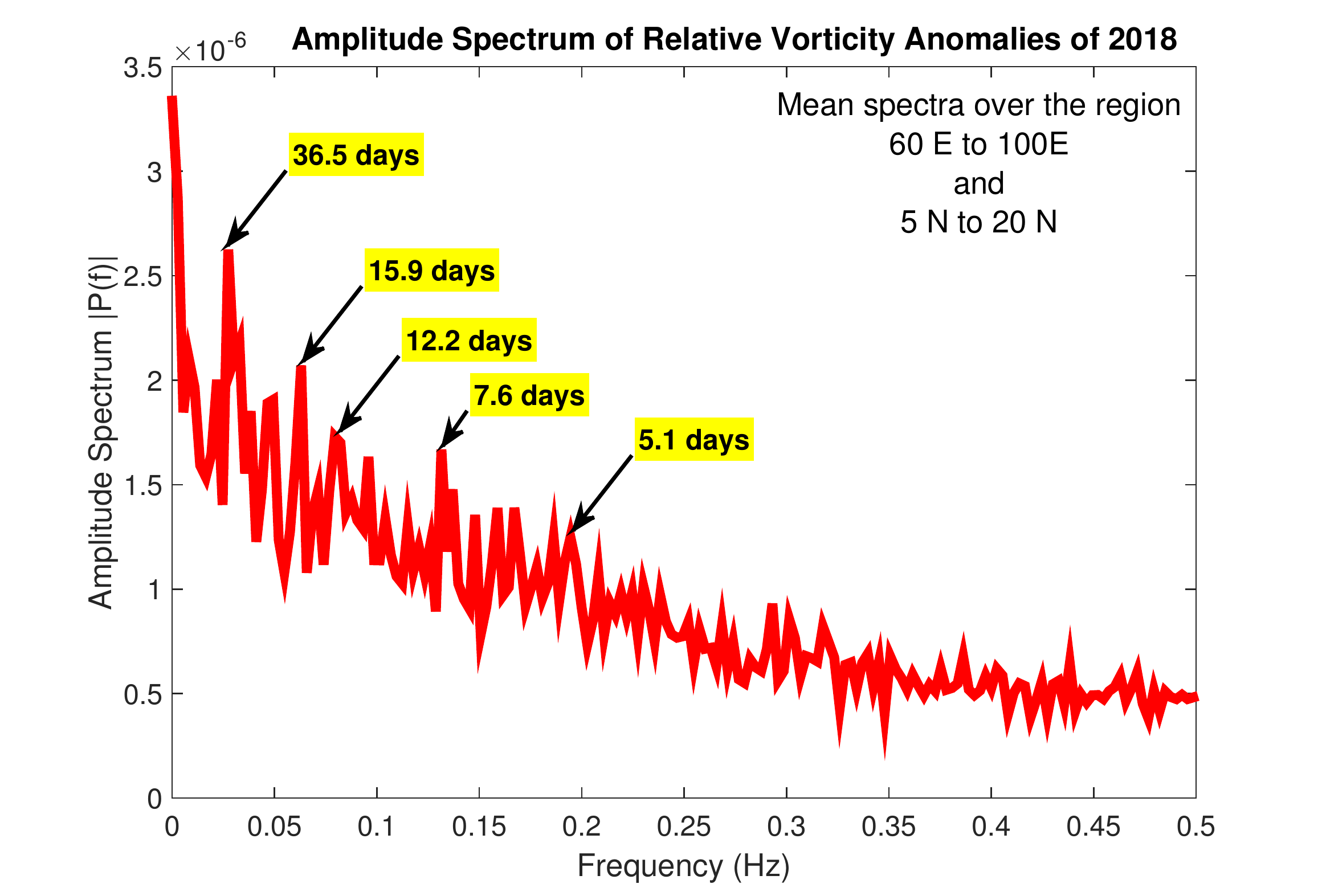}
	\caption{Amplitude spectrum of NCEP/NCAR Reanalysis Relative vorticity Anomalies (in $s^{-1}$) over equatorial Indian Ocean region (averaged over $60^{o}$E to $100^{o}$E \& $5^{o}$N to $20^{o}$N) at 700hPa level in 2018.}
	\label{fig:fig9}
\end{figure}

\begin{figure}
	\centering
	\hspace*{-1cm}
	\includegraphics[width=1.1\linewidth]{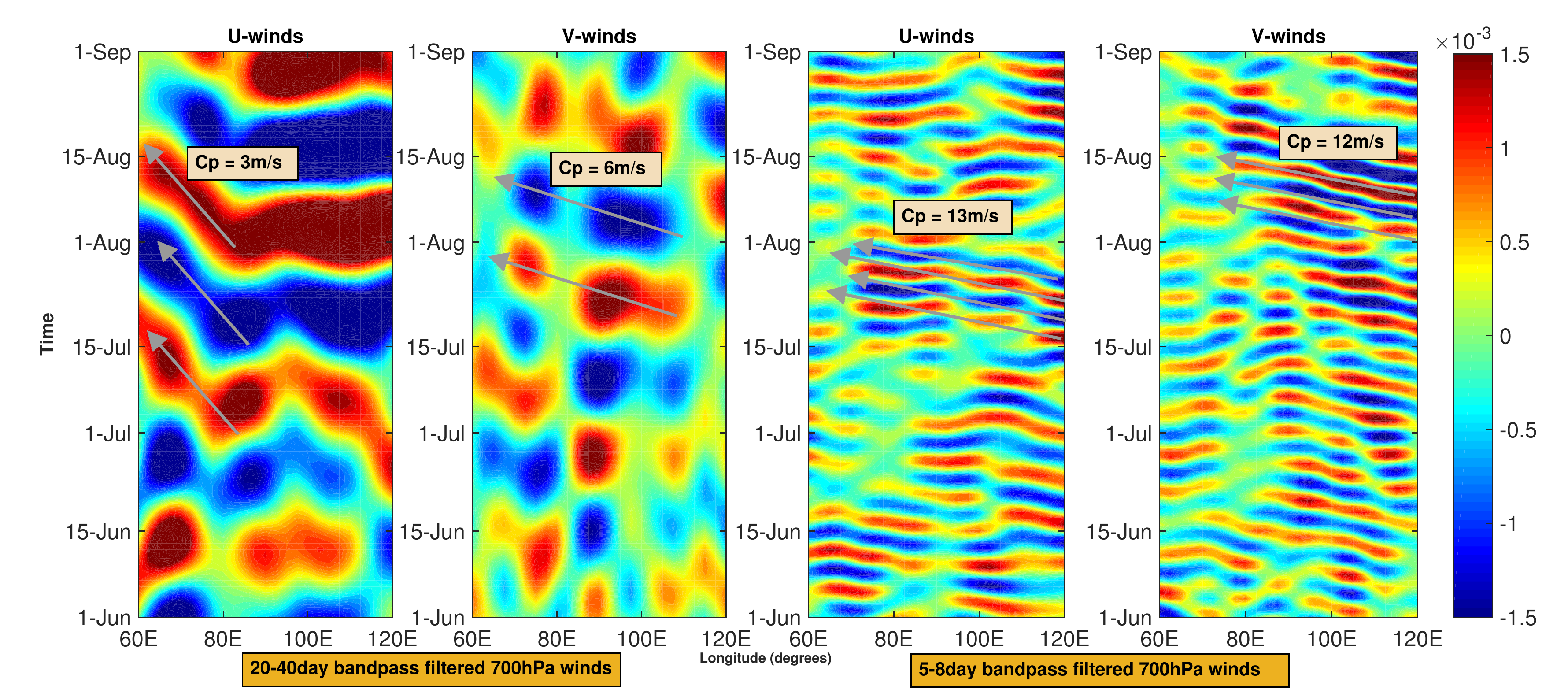}
	\caption{(first \& second figures) 20 to 40 days bandpass-filtered NCEP/NCAR Reanalysis Zonal \& Meridional winds (in m/s) at 700hPa level averaged over the latitudes $5^{o}$N to $15^{o}$N in 2018. (third \& fourth figures) 5 to 8 days bandpass-filtered NCEP/NCAR Reanalysis Zonal \& Meridional winds (in m/s) at 700hPa level averaged over the latitudes $5^{o}$N to $15^{o}$N in 2018. Grey colored arrows indicate propagation of phase in the westward direction. $C_{p}$ represents the phase speed of waves.}
	\label{fig:fig10}
\end{figure}

The presence of prominent periodicities in the wind field may be determined by Fourier Transforms, which converts time domain into frequency domain, as discussed under Section \ref{sec:dataandmethodology}. Fast Fourier Transform (FFT) of the relative vorticity anomalies of 2018 for the region $60^{o}$E to $100^{o}$E \& $5^{o}$N to $20^{o}$N is performed and the spatially averaged resultant amplitude spectrum is given in Figure \ref{fig:fig9}. It may be observed that the dominant periods of variability associated with relative vorticity belong to the ranges such as 5-8 days, 12-16 days and 20-40 days, in addition to the semiannual variability (180 days period corresponds to the much acknowledged strong and sustained westerly winds over the Equatorial Indian Ocean region during Spring and Fall \citep{wyrtki1973}). Hence, the winds at 700 hPa pressure level are bandpass filtered for the above ranges of periods, for further analysis (Fig. \ref{fig:fig10}). Out of the above harmonics, 12-16 days variability is seldom discussed here, owing to its weak contribution to MRG wave dynamics. Therefore, the analysis of the horizontal wind field at the equatorial Indian Ocean and West Pacific (over $60^{o}$E to $120^{o}$E belt, averaged about latitudinal band $5^{o}$N to $15^{o}$N) at 700 hPa, filtered separately for the bands 20-40 days and 5-8 days, are discussed here.

\subsubsection{20-40 day mode} 
From the Hovemöller diagram (Fig. \ref{fig:fig10} (first \& second figures)), one observes westward propagating phase between $60^{o}$E and $90^{o}$E in both zonal and meridional winds at 700 hPa level, with its amplitude maximized during T3 and T4. The waves originate at the eastern end of the equatorial Indian Ocean as a result of intensification of winds over the West Pacific region ($100^{o}$E to $120^{o}$E) perturbed by synoptic scale disturbances, as discussed under Section \ref{subsubsec:mrg}. Hence, the Mixed Rossby-Gravity (MRG) waves excited in the Indian Ocean exist as a dominant mode of 20-40 days periodicity. These MRG waves possess a westward phase speed of 3 m/s in zonal winds and 6 m/s in meridional winds. These disturbances tend to propagate slowly in eastward direction at about 1 m/s. It is interesting to note that the Madden–Julian oscillation (MJO), a prominent convective disturbances in the tropics, exist with the same time period. Moreover, \citet{yang2011} explained that MJO is a wave packet formed as a narrow frequency band of MRG waves, and MJO propagates eastward with the group velocity of MRG waves. Hence, MJO manifested as slow-moving MRG waves of 20-40 day variability in the equatorial Indian Ocean, especially during T3 and T4.

\subsubsection{5-8 day mode}
From the Hovemöller diagram (Fig. \ref{fig:fig10} (third \& fourth figures)), MRG waves of shorter period (5-8 days) is also observed to exist over the Equatorial Indian Ocean. This secondary mode of MRG waves are faster (phase speed of around 12 m/s) but of relatively lower power density, compared to the primary mode of MRG waves (20-40 days). The filtered zonal winds expose the secondary wave mode during T3, while the meridional winds reveal the same during T4. Hence, MRG waves existed in two significant modes in the tropical atmosphere during Indian Monsoon of 2018 - a slow but dominant wave mode of 20-40 days period and a fast mode of 5-8 days period.

\begin{figure}
	\centering
	\hspace*{-1cm}
	\includegraphics[width=1.1\linewidth]{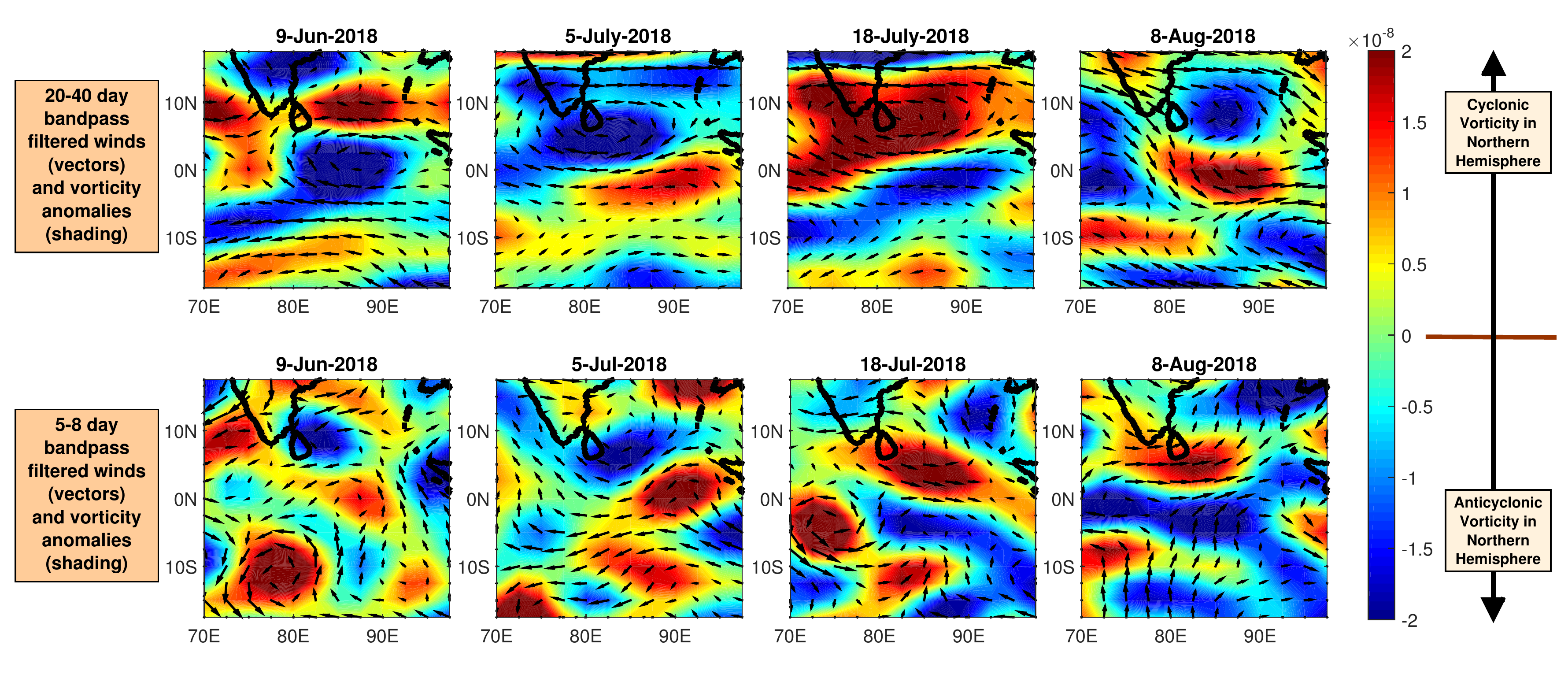}
	\caption{(top) Spatial distribution of 20 to 40 days bandpass-filtered NCEP/NCAR Reanalysis Relative vorticity Anomalies (in $s^{-1}$) at 700hPa level on the days of peak rainfall during T1, T2, T3 and T4. (bottom) Spatial distribution of 5 to 8 days bandpass-filtered NCEP/NCAR Reanalysis Relative vorticity Anomalies (in $s^{-1}$) at 700hPa level for the same days. Positive (negative) values indicate cyclonic (anti-cyclonic) vorticity in the Northern Hemisphere.}
	\label{fig:fig11}
\end{figure}

\begin{figure}
	\centering
	\hspace*{-1cm}
	\includegraphics[width=1.1\linewidth]{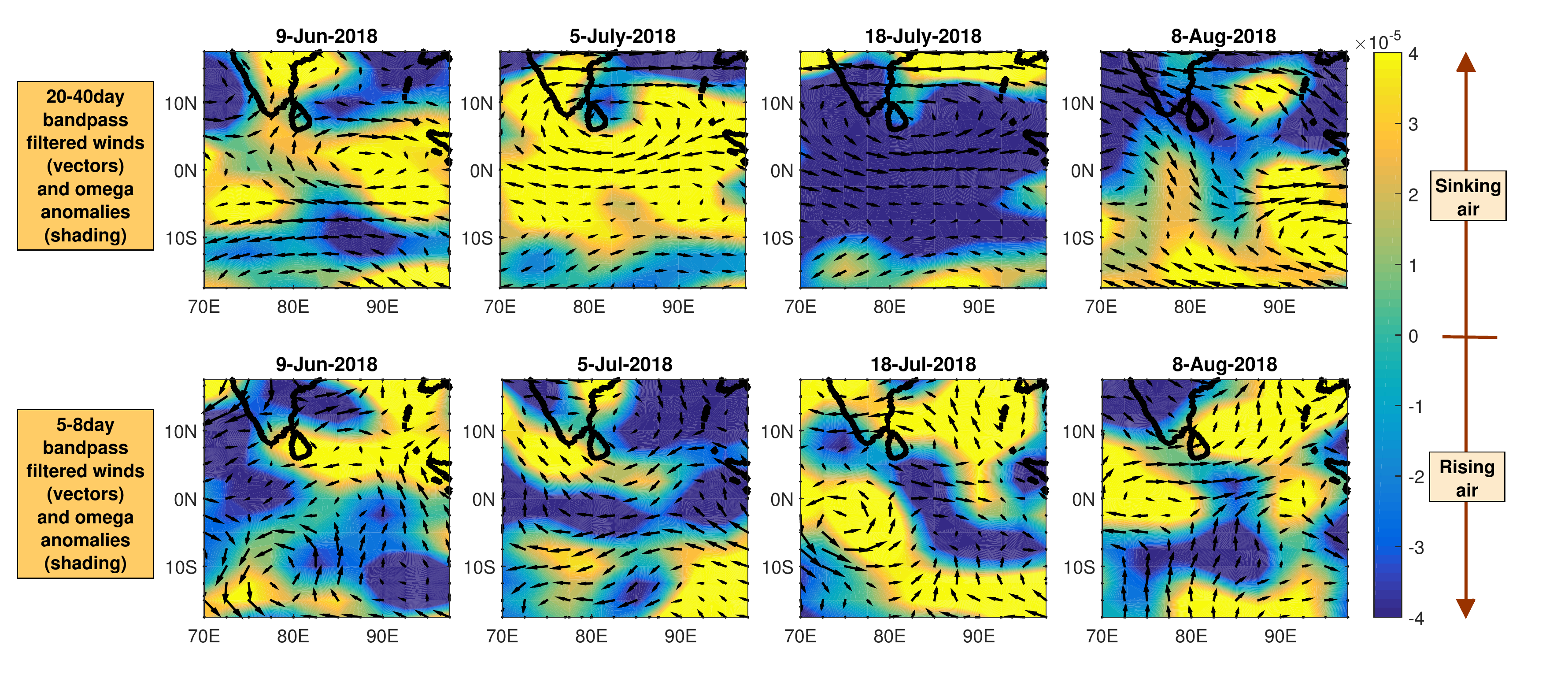}
	\caption{(top) Spatial distribution of 20 to 40 days bandpass-filtered NCEP/NCAR Reanalysis Omega Anomalies (in Pa/s) at 700hPa level on the days of peak rainfall during T1, T2, T3 and T4. (bottom) Spatial distribution of 5 to 8 days bandpass-filtered NCEP/NCAR Reanalysis Omega Anomalies (in Pa/s) at 700hPa level for the same days. Negative (positive) values represent rising (sinking) of air parcels.}
	\label{fig:fig12}
\end{figure}

The above discussion emphasize the prominence of Mixed Rossby-Gravity (MRG) waves, which manifested as dominant modes of variability over Equatorial Indian Ocean in 2018, during the periods of anomalous rainfall over the State of Kerala. These equatorially-trapped waves are convectively-coupled and possess a characteristic wave structure; these waves manifest as synoptic scale gyres centered over as well as symmetric about the equator, but with anti-symmetric convective tendencies \citep{kiladis2009}. Spatial distribution of relative vorticity filtered for periods 20-40 days and 5-8 days each (Fig. \ref{fig:fig11}) show symmetric gyres or vortices positioned over equator, on the days corresponding to peak rainfall during T1, T2, T3 and T4. It may be observed that the circulation remained symmetrical between $15^{o}$N and $15^{o}$S. For example, on $18^{th}$ July 2018 (during T3), a gyre of clockwise circulation (shown in blue shade; it has anticyclonic vorticity in the Northern Hemisphere and cyclonic vorticity in the Southern Hemisphere and represents a wave crest) occupied the equator symmetrically at $85^{o}$E accompanied by a cyclonic vortex (wave trough) over the Indian peninsula and an anti-cyclonic vortex (wave trough) over Southern Indian Ocean. Similarly oriented set of crest and troughs of the MRG wave may be observed on $8^{th}$ August 2018 (during T4) at $75^{o}$E. The characteristic length scale of these vortex structures differ for each modes of MRG waves. The primary mode (20-40 days) of MRG wave has a significant influence over the wind field of 700 hPa level compared to the secondary mode (5-8 days). As explained by \citet{kiladis2009}, MRG wave modes are convectively-coupled and dilate the wind field to favour rising and sinking of air parcels. This shall be investigated with the help of Figure \ref{fig:fig12}, which gives the spatial distribution of bandpass-filtered omega anomalies at 700 hPa level for each MRG wave modes, and shall now aid to assess the convective tendencies forced by MRG wave modes during the days of peak anomalous rainfall. Figure \ref{fig:fig12} (top) reveals that the primary mode brought the entire State of Kerala under a deep convective zone (high negative omega anomalies) during T1, T3 and T4, while the secondary mode caused convection only to the northern part of the State during T2. Upon comparison of spatial distributed omega anomalies (Fig. \ref{fig:fig12}) with that of the precipitation (Fig. \ref{fig:fig2}), it may be concluded that for the primary mode of MRG waves, the omega anomalies bears high spatial correlation with the rainfall on the days of peak anomalies in precipitation. Hence, the MRG wave mode of 20-40 days period had a very profound impact on the rainfall across Kerala in 2018. 
Yet, the faster MRG wave (secondary mode) cannot be neglected, as it assumed significance during T2, when only the northern part of the State fell within the convective zone. Therefore, both the primary and secondary modes of MRG waves in superposition, played a pivotal role in driving convection during the period of anomalous rainfall in the State of Kerala in 2018.

\section{Summary \& Conclusions}
The State of Kerala experienced a spate of flood events during the South-West Monsoon of 2018. It is much more than a hydrogeological problem, especially in the pretext of growing concerns of changing climate and its serious ramifications. Hence, this study specifically unearths the dynamical phenomenon responsible for the anomalous course of events.

Kerala received very high precipitation in 2018, especially during the fag end of July (T3) and beginning of August (T4). The northern part of the state received more anomalous rainfall, with daily precipitation departures as high as 150 mm on $8^{th}$ August. Western Ghats played a major role to limit the spatial extend of extreme precipitation to its windward side. The Intertropical Convergence Zone (ITCZ) cloud band maxima shifted a large number of times to the mainland and centered over the State, but underwent substantial fluctuations in its position during T3 and T4. Besides, the ITCZ widened in tune with high meridional rainfall maxima during periods of anomalous precipitation. This interesting, yet unprecedented behaviour of Equatorial trough motivates one to foray into the atmospheric dynamics which perturbed the tropical convective system.

The synoptic-scale disturbances of the Pacific or Atlantic Ocean travelled westward and perturbed the tropical dynamics to cause heavy downpour over Kerala in 2018. They triggered the equatorially-trapped Mixed Rossby-Gravity (MRG) Waves over the eastern Equatorial Indian Ocean at 700 hPa pressure level, and propagated westward with an average phase speed of 10 m/s to the strike coast of Africa. They possessed an eastward group velocity of around 6 m/s. These waves tend to escalate the relative vorticity as they propagate through the medium and meander the wind field. The MRG waves perturbed the ITCZ too; the westerly wave group caused the convective trough to widen. Further, their phase propagates vertically upward to the upper troposphere accompanied by intense updraught of air, thereby fueling convection in the tropics. 

The Equatorial Indian Ocean basin resonated to the two convectively-coupled modes of MRG waves -- a primary slow-moving wave of period 20-40 days and a secondary faster wave of period 5-8 days. The slow westward propagating wave (phase speed of 3-6 m/s) has the period identical to that of Madden-Julian Oscillations (MJO), which is a characteristic MRG wave group travelling eastward. The westward propagating faster waves are characteristic MRG waves with phase speed of around 12-13 m/s. Both the wave modes set up vortices (both cyclonic and anti-cyclonic) in the medium and are hence responsible for meandering of winds. These circulations or atmospheric gyres position symmetrically over the equator, as they propagate across the basin and dilate the wind field to establish zones of convection in the tropics. The primary mode of MRG waves (20-40 days period) is mainly responsible for stimulated convection which resulted into large anomalies in rainfall over Kerala in 2018. The faster secondary mode, although of relatively low power density, could significantly describe the spatial distribution of convective zones during first week of July. Therefore, tropical atmosphere resonated to the two MRG wave modes in superposition and further, catalyzed convection over the equatorial Indian Ocean, which resulted in heavy precipitation over the State of Kerala in 2018.

\bibliographystyle{unsrtnat}
\bibliography{references}  

\newpage

\begin{table}
	\caption{Rainfall at Pavannur \& Mankadavu in Kerala during the period of anomalous rainfall in 2018}
	\centering
	\begin{tabular}{llllll}
		\toprule
		\multicolumn{2}{c}{Period of Anomalous Rainfall} & \multicolumn{2}{c}{Pavannur Rainfall (mm/day)} & \multicolumn{2}{c}{Mankadavu Rainfall (mm/day)}\\ 
		\cmidrule(r){1-2}
		\cmidrule(r){3-4}
		\cmidrule(r){5-6}
		Notation  & Period  & Max. & Anomaly & Max. & Anomaly \\
		\midrule
		T1 & 08-June-2018 to 11-June-2018  & 98.50 & 47.41 & 69.20 & 51.94 \\
		& (Peak Rainfall on 09-June-2018) \\
		\vspace{0.1cm}\\
		T2 & 03-July-2018 to 07-July-2018 & 105.70 & 90.06 & 7.27 & 4.30 \\
		& (Peak Rainfall on 05-July-2018) \\
		\vspace{0.1cm}\\
		T3 & 16-July-2018 to 26-July-2018 & 196.50 & 148.80 & 49.52 & 11.01 \\
		& (Peak Rainfall on 18-July-2018) \\
		\vspace{0.1cm}\\
		T4 & 04-August-2018 to 12-August-2018  & 249.50 & 149.30 & 97.26 & 54.81 \\
		& (Peak Rainfall on 08-August-2018) \\
		\bottomrule
	\end{tabular}
	\label{tab:table1}
\end{table}








\end{document}